\documentclass[11pt]{article}

\usepackage{amssymb}
\usepackage{graphics}

\newtheorem{thm}{Theorem}
\newtheorem{claim}{Claim}[thm]

\def \no {\noindent}

\title{Weighted Efficient Domination in  \\Classes of $P_6$-free Graphs}

\author{Andreas Brandst\"adt\thanks{Institut f\"ur Informatik, Universit\"at Rostock, D-18051 Rostock, Germany.
E-mail: {andreas.brandstaedt@uni-rostock.de}} \and T. Karthick\thanks{Corresponding author. Computer Science Unit, Indian Statistical
Institute, Chennai Centre, Chennai-600113, India.
E-mail: {karthick@isichennai.res.in}}}

\begin{document}
\maketitle

\begin{abstract} In a graph $G$, an {\it efficient dominating set} is a subset $D$ of vertices such that $D$ is an
independent set and each vertex outside $D$ has exactly one neighbor in $D$.
The {\textsc{Minimum Weight Efficient Dominating Set (Min-WED)}} problem   asks for an efficient dominating set of total minimum weight in a given vertex-weighted graph; the {\textsc{Maximum Weight Efficient Dominating Set  (Max-WED)}} problem is defined similarly.
The  {\textsc{Min-WED/Max-WED}} is known to be $NP$-complete for $P_7$-free graphs, and is known to be polynomial time solvable for $P_5$-free graphs.
However, the computational complexity of the  {\textsc{min-WED/max-WED}} problem is unknown for $P_6$-free graphs.
In this paper, we show that the {\textsc{Min-WED/Max-WED}} problem  can be solved in polynomial time for two subclasses of $P_6$-free graphs, namely for
 ($P_6,S_{1,1,3}$)-free graphs, and for ($P_6$, bull)-free graphs.
\end{abstract}

\no{\bf Keywords}: Graph algorithms; Domination in graphs; Efficient domination; Perfect code; $P_6$-free graphs.

\section{Introduction}

Throughout this paper, let $G = (V, E)$ be a finite, undirected and simple graph  with $n$ vertices and $m$ edges.  For notation and terminology not defined here, we follow \cite{BraLeSpi1999}.
In a graph $G$, a subset $D \subseteq V$ is a {\it dominating set} if each vertex outside $D$ has some neighbor in $D$. An
{\it efficient dominating set} ({\em e.d.}) is a dominating set $D$ such that $D$ is an independent set and each vertex outside $D$ has exactly one neighbor in $D$.
Efficient dominating sets were introduced by Biggs \cite{Biggs1973}, and are also called perfect codes, perfect dominating sets and independent perfect dominating sets in the literature. The notion of efficient dominating sets is motivated by various interesting applications such as coding theory and resource allocation in parallel computer networks; see \cite{Biggs1973,LivSto1988}. We refer to \cite{HayHedSla1998} for more information on efficient domination in graphs.

\medskip

The {\textsc{Efficient Dominating Set (ED)}} problem asks for the existence of an efficient dominating set in a given graph $G$.
The {\textsc{Minimum Weight Efficient Dominating Set (Min-WED)}} problem asks for an efficient dominating set of total minimum weight in a given vertex-weighted graph; the {\textsc{Maximum Weight Efficient Dominating Set (Max-WED)}} problem is defined similarly.

Clearly, a graph $G =(V,E)$ has an efficient dominating set if and only if $(G, w, |V|)$ is a yes instance to the {\textsc{Min-WED}} problem, where $w(v) =1$, for every $v \in V$, and the  {\textsc{Min-WED} problem is equivalent to the {\textsc{Max-WED}} problem (see \cite{BraFicLeiMil2015}).

\medskip

The ED problem is known to be $NP$-complete in general, and is known to be $NP$-complete for several restricted classes of graphs such as: bipartite graphs \cite{YenLee1996}, chordal graphs \cite{YenLee1996}, chordal bipartite graphs \cite{LuTan2002}, planar bipartite graphs \cite{LuTan2002}, and planar graphs with maximum degree three \cite{FelHoo1991}. However, ED is solvable in polynomial time for split graphs \cite{ChaLiu1993}, co-comparability graphs \cite{Chang1997, ChaPanCoo1995}, interval graphs \cite{ChaLiu1994}, circular-arc graphs \cite{ChaLiu1994}, and for many more classes of graphs (see
\cite{BraFicLeiMil2015,BraMilNev2013} and the references therein).

\medskip

Let $P_k$ denote the chordless path with $k$ vertices and let $C_k$ denote the chordless cycle with $k$ vertices, $k \ge 3$. A \emph{hole} is a chordless cycle $C_k$, where $k \geq 5$.
Let $S_{i, j, k}$ denote a tree with exactly three vertices of degree one, being at distance $i$, $j$ and $k$ from the unique vertex of degree three.  Note that $S_{i,j,0}$ is a path on $i +j +1$ vertices, while $S_{1,1,1}$ is called a {\it claw} and $S_{1, 1, 2}$ is called a {\it chair or fork}.
See Figure \ref{SBB} for some special graphs used in this paper.

\medskip

In this paper, we focus on the {\textsc{Min-WED/Max-WED}} problem in certain classes of graphs that are defined by forbidden induced subgraphs. If $\cal{F}$ is a family of graphs, a graph $G$ is said to be {\it $\cal{F}$-free} if it does not contain any induced subgraph isomorphic to any graph in $\cal{F}$.
The ED problem is known to be $NP$-complete for ($K_{1,3}, K_4-e$)-free perfect graphs \cite{LuTan1998}, and for $2P_3$-free chordal graphs \cite{SmaSla1995}. In particular, ED is $NP$-complete for $P_7$-free graphs.

\begin{figure}[t]
\centering
 \includegraphics{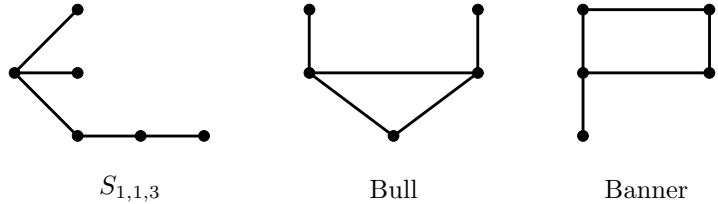}
\caption{Some special graphs.}
\label{SBB}
\end{figure}

\medskip

Recently, Brandst\"adt et al. \cite{BraMilNev2013} gave a linear time algorithm for solving the  {\textsc{Min-WED/Max-WED}} on $2K_2$-free graphs, and showed that the  {\textsc{Min-WED/Max-WED}} is solvable in polynomial time for $P_5$-free graphs. Brandst\"adt and Le \cite{BraLe2014} showed that the  {\textsc{Min-WED/Max-WED}} is solvable in polynomial time for (E, xNet)-free graphs, thereby extending the result on $P_5$-free graphs. However, the computational complexity of ED is unknown for $P_6$-free graphs. Brandst\"adt et al. showed that ED is solvable in polynomial time for ($P_6, S_{1,2,2}$)-free graphs \cite{BraMilNev2013}, ($P_6$, HHD)-free graphs, and ($P_6$, house)-free graphs \cite{BraEscFri2015}. It has also been shown that the {\textsc{Min-WED/Max-WED}} can be solved in polynomial time for ($P_6$, banner)-free graphs \cite{Karth2015}. We refer to Figure 1 of \cite{BraFicLeiMil2015, BraMilNev2013} for the complexity of ED {\textsc{Min-WED/Max-WED}} on several graph classes.

\medskip

For a graph $G=(V,E)$ and two vertices $u,v \in V$, let $d_G(u,v)$ denote the {\it distance} between $u$ and $v$ in $G$. The {\it square} of $G$ is the graph
$G^2 = (V, E^2)$ such that $uv \in E^2$ if and only if $d_G(u,v) \in \{1, 2\}$.

\medskip
In an undirected graph $G$, an {\it independent set} is a set of mutually nonadjacent vertices.
 The \textsc{Maximum Weight Independent Set (MWIS)} problem
asks for an independent set of maximum total  weight in the given graph $G$ with vertex weight function $w$ on  $V(G)$.
Recently, Brandst\"adt et al. \cite{BraFicLeiMil2015} developed a framework for solving the weighted efficient domination problems based on a reduction to the MWIS problem in the square of the input graph, and is given below.

\begin{thm}[\cite{BraFicLeiMil2015}]\label{EDS-MWIS-reduction}
Let $\cal{C}$ be a graph class for which the MWIS problem is solvable in time $T(|G|)$ on squares of graphs from $\cal{C}$.
Then the {\textsc{Min-WED/Max-WED}} problems are solvable on graphs in $\cal{C}$ in time $O(\min\{nm+n, n^{\mu}\}+T(|G^2|))$, where $\mu < 2.3727$ is the matrix multiplication exponent \cite{Willi2012}.
\end{thm}


In this paper, using the above framework, we show that the {\textsc{Min-WED/ Max-WED}} problem can be solved  in polynomial time in two subclasses of $P_6$-free graphs, namely ($P_6, S_{1,1,3}$)-free graphs and ($P_6$, bull)-free graphs. In particular, we prove the following:
\begin{enumerate}
\item[(1)] If $G$ is a ($P_6,S_{1,1,3}$)-free graph that has an efficient dominating set, then $G^2$ is $P_5$-free (Section 2, Theorem \ref{P6-S113-free-implies-P5-free}).
\item[(2)] If $G$ is a ($P_6$,bull)-free graph that has an efficient dominating set, then $G^2$ is (hole,banner)-free (Section 3,  Theorems \ref{GP6bullfredG2holefr} and \ref{P6-bull-free-ed-implies-banner-free}).
\end{enumerate}

Since MWIS can be solved in polynomial time for $P_5$-free graphs \cite{LokVatVil2014} and for (hole,banner)-free graphs (Section 3, Theorem \ref{hole-banner-mwis-time}), our results follow from (1), (2)  and Theorem \ref{EDS-MWIS-reduction}.

\medskip
Note that the class of $P_5$-free graphs is a subclass of ($P_6, S_{1,1,3}$)-free graphs. Also, note that from the $NP$-completeness result for $K_{1,3}$-free graphs \cite{LuTan1998}, it follows that for $S_{1,1,3}$-free graphs, ED remains $NP$-complete. The class of bull-free graphs includes some well studied classes of graphs in the literature such as: $P_4$-free graphs, triangle-free graphs, and paw-free graphs.

\section{Weighted Efficient Domination in ($P_6,S_{1,1,3}$)-free graphs}

In this section, we show that the {\textsc{Min-WED/Max-WED}} can be solved efficiently in ($P_6,S_{1,1,3}$)-free graphs. First, we prove the following:

\begin{thm}\label{P6-S113-free-implies-P5-free}
Let $G = (V,E)$ be a $(P_6,S_{1,1,3})$-free graph. If $G$ has an efficient dominating set, then $G^2$ is $P_5$-free.
\end{thm}

\no{\bf Proof.}
Let $G$ be a ($P_6,S_{1,1,3}$)-free graph having an efficient dominating set $D$, and suppose to the contrary that $G^2$ contains an induced $P_5$, say with vertices $v_1, \ldots, v_5$ and edges $v_iv_{i+1} \in E^2$, $i \in \{1,2,3,4\}$.  Then $d_G(v_i,v_{i+1}) \leq 2$ for $i \in \{1,2,3,4\}$ while $d_G(v_i,v_j) \geq 3$ for $|i-j| \ge 2$.

 We can assume that $d_G(v_i,v_{i+1}) = 2$ for all $i \in \{1,2,3,4\}$ since in all other cases it is easily verified that either $P_6$ or $S_{1,1,3}$ is an induced subgraph of~$G$.
For $i \in \{1,2,3,4\}$, let $x_i$ be a common neighbor of $v_i$ and $v_{i+1}$.
Note that, by the distance properties, there are no other edges between the vertex sets  $\{x_1,x_2,x_3,x_4\}$ and $\{v_1,v_2,v_3,v_4,v_5\}$.

\begin{claim}\label{compedge-1}
$x_1x_4 \in E$.
\end{claim}

\no{\it Proof of Claim $\ref{compedge-1}$}: Suppose to the contrary that $x_1x_4 \notin E$. We claim that this implies $x_1x_3 \notin E$ and $x_2x_4 \notin E$:
Suppose that $x_1x_3 \in E$. Then, if $x_3x_4 \in E$, $\{v_5,x_4,x_3,x_1,v_1,v_2\}$ induces an $S_{1,1,3}$ in $G$, and if $x_3x_4 \notin E$, $\{v_5,x_4,v_4,x_3,x_1,v_1\}$ induces a $P_6$ in $G$, which is a contradiction. Thus, under the assumption that $x_1x_4 \notin E$, we have $x_1x_3 \notin E$, and by symmetry, we have $x_2x_4 \notin E$.

Thus, if $x_1x_4 \notin E$, the only possible edges among $\{x_1,x_2,x_3,x_4\}$ are the pairs $x_ix_{i+1}$, $1 \le i \le 3$, but if all three are edges then $\{v_1,x_1,x_2,x_3,x_4,v_5\}$ induces a $P_6$ in $G$, and if at least one of the pairs $x_ix_{i+1}$ is a non-edge, we have an induced $P_6$ in each case, which is a contradiction.

\no Hence Claim \ref{compedge-1} holds.
$\Diamond$

\begin{claim}\label{compedge-2}
$x_1x_3 \in E$ and $x_2x_4 \in E$.
\end{claim}

\no{\it Proof of Claim $\ref{compedge-2}$}: Without loss of generality, suppose to the contrary that $x_2x_4 \notin E$.
Then $x_1x_2 \in E$ (otherwise, $\{x_2,v_2,x_1,x_4,v_4,v_5\}$ induces an $S_{1,1,3}$ in $G$) but then $\{v_3, x_2, x_1, x_4, v_4, v_5\}$ induces an  $S_{1,1,3}$ in $G$, which is a contradiction. Thus, $x_2x_4 \in E$ and by symmetry, also $x_1x_3  \in E$ holds.

\no Hence Claim \ref{compedge-2} is shown.
$\Diamond$

\medskip

By Claims \ref{compedge-1} and \ref{compedge-2}, we have $x_1x_3, x_1x_4, x_2x_4 \in E$. Our next step is:

\begin{claim}\label{abcd-notin-D}
For all $i \in \{1,2,3,4\}$, $x_i \notin D$.
\end{claim}

\no{\it Proof of Claim $\ref{abcd-notin-D}$}:

(i) First, suppose to the contrary that $x_1 \in D$. Then, since $D$ is an e.d., $v_4, x_4, v_5 \notin D$. So, there exists $v_5' \in D$ such that $v_5v_5' \in E$. Since $D$ is an e.d., $v_5'x_4 \notin E$ and $v_5'x_1 \notin E$, and by the distance properties, $v_5'v_1 \notin E$ and $v_5'v_2 \notin E$.
Thus, $\{v_5',v_5,x_4,x_1,v_1,v_2\}$ induces an $S_{1,1,3}$ in $G$ which is a contradiction showing that $x_1 \notin D$.
By symmetry, we obtain $x_4 \notin D$.

(ii) Now, suppose to the contrary that $x_2 \in D$. Then, since $D$ is an e.d., $v_4, x_4, v_5 \notin D$. So, there exists $v_5' \in D$ such that $v_5v_5' \in E$. Since $D$ is an e.d., $v_5'x_4 \notin E$ and $v_5'x_2 \notin E$, and by the distance properties, $v_5'v_2 \notin E$ and $v_5'v_3 \notin E$.
Now $\{v_5', v_5,x_4,x_2,v_2,v_3\}$ induces an $S_{1,1,3}$ in $G$, which is a contradiction. Hence, $x_2 \notin D$.
By symmetry, we obtain $x_3 \notin D$.

\no Hence Claim \ref{abcd-notin-D} holds.
$\Diamond$

\begin{claim}\label{v2v4-notin-D}
$v_2 \notin D$ and $v_4 \notin D$.
\end{claim}

\no{\it Proof of Claim $\ref{v2v4-notin-D}$}: Without loss of generality, suppose to the contrary that $v_2 \in D$.
If $v_4 \in D$ then, since $D$ is an e.d., $v_5 \notin D$. So, there exists $v_5' \in D$ such that $v_5v_5' \in E$. Again, since $D$ is an e.d., $v_5'$ is not adjacent to $x_1,x_4,v_2$, and by the distance properties, $v_5'v_1 \notin E$. Thus, $\{v_5', v_5, x_4, x_1, v_1, v_2\}$ induces an $S_{1,1,3}$ in $G$, which is a contradiction.

Hence $v_4 \notin D$ holds. Since by the distance properties, $v_2v_4 \notin E$ and by Claim \ref{abcd-notin-D}, $x_3,x_4 \notin D$, there exists $v_4' \in D$ such that $v_4v_4' \in E$. Now, if $v_4'x_4 \notin E$, then since $v_4'x_1 \notin E$ and $v_4'v_2 \notin E$ since $D$ is an e.d., and since by the distance properties, $v_4'v_1 \notin E$, $\{v_4', v_4, x_4, x_1, v_1, v_2\}$ induces an $S_{1,1,3}$ in $G$, which is a contradiction. Thus, $v_4'x_4 \in E$ holds, and by a similar argument, $v_4'x_3 \in E$ also holds. This implies $v_5 \notin D$ since $D$ is an e.d. Hence there exists $v_5' \in D$ such that $v_5v_5' \in E$. Since $\{v_5, v_4', x_3, x_1, v_1, v_2\}$ does not induce an $S_{1, 1, 3}$ in $G$, we have $v_5' \neq v_4'$.  Then, since $D$ is an e.d., $v_5'x_4 \notin E$,  $v_5'x_1 \notin E$ and
 $v_5'v_2 \notin E$, and by the distance properties $v_5'v_1 \notin E$. Now, $\{v_5', v_5, x_4, x_1, v_1, v_2\}$ induces an $S_{1,1,3}$ in $G$, which is a contradiction.

\no Hence Claim \ref{v2v4-notin-D} is shown.
$\Diamond$

\medskip

Since $x_1, x_2, v_2 \notin D$ (by Claims \ref{abcd-notin-D} and \ref{v2v4-notin-D}), there exists $v_2' \in D$ such that $v_2v_2' \in E$. Moreover, since $x_3, x_4, v_4 \notin D$ (by Claims \ref{abcd-notin-D} and \ref{v2v4-notin-D}), there exists  $v_4' \in D$ such that $v_4v_4' \in E$. Note that by the distance properties, we have $v_2' \neq v_4'$. Then we prove the following:

\begin{claim}\label{v1v5-notin-D}
$v_1 \notin D$ and $v_5 \notin D$.
\end{claim}

\no{\it Proof of Claim $\ref{v1v5-notin-D}$}: Without loss of generality, suppose to the contrary that $v_1 \in D$. Then, since $D$ is an e.d.,
$v_4'x_1 \notin E$, and by the distance properties, $v_4'v_1 \notin E$ and $v_4'v_2 \notin E$. This implies $v_4'x_4 \in E$ since $\{v_4',v_4,x_4,x_1,v_1,v_2\}$ does not induce an $S_{1,1,3}$ in $G$. Since $D$ is an e.d., $v_2'x_1 \notin E$ and $v_2'x_4 \notin E$, and by the distance properties, $v_2'v_4 \notin E$ and $v_2'v_5 \notin E$ but then $\{v_2', v_2, x_1, x_4, v_4, v_5\}$ induces an $S_{1,1,3}$ in $G$, which is a contradiction.
A symmetric argument shows that $v_5 \notin D$.

\no Hence Claim \ref{v1v5-notin-D} holds.
$\Diamond$

\begin{claim}\label{av4'-dv2'-in-E}
$v_4'x_1 \in E$ and $v_2'x_4 \in E$.
\end{claim}

\no{\it Proof of Claim $\ref{av4'-dv2'-in-E}$}: Without loss of generality, suppose to the contrary that $v_4'x_1 \notin E$. By the distance properties, $v_4'$ is not adjacent to $v_1$ and $v_2$. We first claim that $v_4'x_3 \in E$ and $v_4'x_4 \in E$:

If $v_4'x_3 \notin E$ then $\{v_4', v_4, x_3, x_1, v_1, v_2\}$ induces an $S_{1,1,3}$ in $G$, and if $v_4'x_4 \notin E$ then $\{v_4', v_4, x_4, x_1, v_1, v_2\}$ induces an $S_{1,1,3}$ in $G$, which is a contradiction. Thus, $v_4'x_3 \in E$ and $v_4'x_4 \in E$ holds.

Since $v_1 \notin D$ (by Claim \ref{v1v5-notin-D}), there exists $v_1' \in D$ such that $v_1v_1' \in E$. Note that by Claim \ref{abcd-notin-D}, $v_1' \neq x_1$.
By the distance properties, $v_1'v_4 \notin E$ and $v_1'v_5 \notin E$, and since $D$ is an e.d., $v_1'x_4 \notin E$.
If $v_1' x_1 \notin E$ then $\{v_1', v_1, x_1, x_4, v_4, v_5\}$ induces an $S_{1,1,3}$ in $G$, which is a contradiction.
Thus $v_1' x_1 \in E$.

We claim that $v_1' \neq v_2'$: If $v_1' = v_2'$ then in the case that $v_2' x_2 \in E$, $\{v_1, v_2', x_2, x_4, v_5, v_4\}$ induces an $S_{1,1,3}$ in $G$, and in
the other case when $v_2' x_2 \notin E$, $\{v_1, v_2', v_2, x_2, x_4, v_5\}$ induces a $P_6$ in $G$, which is a contradiction. Thus, $v_1' \neq v_2'$ holds.

Hence, since $D$ is an e.d., $v_2'x_1 \notin E$ and $v_2'x_4 \notin E$. Also, by the distance properties, $v_2'$ is not adjacent to $v_4$ and $v_5$. Now,
$\{v_2', v_2, x_1, x_4, v_4, v_5\}$ induces an $S_{1,1, 3}$ in $G$, which is a contradiction. This finally shows that $v_4'x_1 \in E$ holds.

By symmetric arguments, we can show $v_2'x_4 \in E$.

\no Hence Claim \ref{av4'-dv2'-in-E} holds.
$\Diamond$

\medskip


Next, we have the following:

\begin{claim}\label{v_1'=v2'+ v4'=v5'}
$v_2'v_1 \in E$ and $v_4'v_5 \in E$.
\end{claim}

\no{\it Proof of Claim $\ref{v_1'=v2'+ v4'=v5'}$}: Without loss of generality, suppose to the contrary that $v_2'v_1 \notin E$.
Since $v_1 \notin D$, there exists $v_1' \in D$  such that $v_1' \neq v_2'$ and $v_1v_1' \in E$. Then, by Claim \ref{av4'-dv2'-in-E} and since $D$ is an e.d.,
we have $v_1'x_1 \notin E$ and $v_1'x_4 \notin E$, and by the distance properties, $v_1'v_4 \notin E$ and $v_1'v_5 \notin E$.
Now, $\{v_1', v_1, x_1, x_4, v_4, v_5\}$ induces an $S_{1,1,3}$ in $G$, which is a contradiction.

By symmetric arguments, we obtain $v_4'v_5 \in E$.

\no Hence Claim \ref{v_1'=v2'+ v4'=v5'} holds.
$\Diamond$

\medskip

Next we have:

\begin{claim}\label{v3-notin-D}
$v_3 \notin D$.
\end{claim}

\no{\it Proof of Claim $\ref{v3-notin-D}$}:
Suppose to the contrary that $v_3 \in D$. Since $D$ is an e.d., this implies that $v_2'x_2 \notin E$, $v_2'v_3 \notin E$, and $v_2'x_3 \notin E$. But then $\{v_1,v_2', v_2, x_2, v_3, x_3, v_4\}$ induces a $P_6$ in $G$, which is a contradiction.
Hence Claim \ref{v3-notin-D} holds.
$\Diamond$

\medskip

Thus, $v_3 \notin D$. By Claim \ref{abcd-notin-D}, $x_2,x_3 \notin D$. Thus, there is $v_3' \in D$ with $v_3v_3' \in E$, and by Claim \ref{v_1'=v2'+ v4'=v5'} and by the distance properties, $v'_3 \neq v'_2$ and $v'_3 \neq v'_4$ holds. Then we have the following.

\begin{claim}\label{v_3'b, v3'c-in-E}
$v_3'x_2 \in E$ and $v_3'x_3 \in E$.
\end{claim}

\no{\it Proof of Claim $\ref{v_3'b, v3'c-in-E}$}: Without loss of generality, suppose to the contrary that $v_3'x_2 \notin E$.
Then, by Claims \ref{av4'-dv2'-in-E} and \ref{v_1'=v2'+ v4'=v5'}, $\{v_3', v_3, x_2, x_4, v_4, v_5\}$ induces an $S_{1,1,3}$ in $G$, which is a contradiction.
Thus, $v_3'x_2 \in E$, and by a symmetric argument, we have $v_3'x_3 \in E$ which shows Claim $\ref{v_3'b, v3'c-in-E}$.
$\Diamond$

\medskip

Now, since $D$ is an e.d., we see that $\{v_1,v_2', v_2, x_2, v_3, x_3, v_4\}$ induces a $P_6$ in $G$, which is a contradiction. This finishes the proof of Theorem \ref{P6-S113-free-implies-P5-free}.
\hfill{$\Box$}

\begin{thm}\label{EDS-P6-S113-free}
The {\textsc{Min-WED/Max-WED}} problem can be solved in polynomial time for $(P_6,S_{1,1,3})$-free graphs.
\end{thm}

\no{\bf Proof.}
Since the MWIS problem in $P_5$-free graphs can be solved in polynomial time \cite{LokVatVil2014}, Theorem \ref{EDS-P6-S113-free} follows by
 Theorems \ref{EDS-MWIS-reduction} and \ref{P6-S113-free-implies-P5-free}.
\hfill{$\Box$}

\section{Weighted Efficient Domination in ($P_6$, bull)-free graphs}

Brandst\"adt et al. \cite{BraEscFri2015} showed that if $G$ is a ($P_6$, bull)-free graph that has an efficient dominating set, then $G^2$ is perfect. Since MWIS can be solved in polynomial time for perfect graphs \cite{GroLovSch1981}, WED can be solved in polynomial time for ($P_6$, bull)-free graphs.

In this section, we show that WED can be solved more efficiently in time $O(n^2 m)$ for ($P_6$, bull)-free graphs (which considerably improves the time bound for this graph class).

\subsection{Squares of ($P_6$, bull)-free graphs with e.d. are hole-free}

In \cite{BraEscFri2015}, the following is shown:
\begin{thm}[\cite{BraEscFri2015}]\label{GP6fredG2holefr}
Let $G = (V, E)$ be a $P_6$-free graph. If $G$ has an efficient dominating set then $G^2$ is hole-free.
\end{thm}

Though it directly follows from Theorem \ref{GP6fredG2holefr} that for any ($P_6$, bull)-free graph $G$ with e.d., its square $G^2$ is hole-free, the structure for ($P_6$, bull)-free graphs is more special; we describe this in Theorem \ref{GP6fredG2holefr} and we give a direct proof for Theorem \ref{GP6bullfredG2holefr} here which is much simpler for the subclass of ($P_6$, bull)-free graphs and makes this paper self-contained.

\begin{thm}\label{GP6bullfredG2holefr}
Let $G=(V,E)$ be a $(P_6$,bull$)$-free graph. Then we have:
\begin{enumerate}
\item[$(i)$] $G^2$ is $C_k$-free for all $k \ge 6$.
\item[$(ii)$] If $G$ has an efficient dominating set then $G^2$ is $C_5$-free.
\end{enumerate}
\end{thm}

\no{\bf Proof.} Let $G$ be a ($P_6$, bull)-free graph, and let $H$ denote a hole (isomorphic to $C_k$, $k \ge 5$) in $G^2$ with vertices $\{v_1, v_2, \ldots, v_k\}$ and edges $v_iv_{i+1} \in E^2$ (index arithmetic modulo $k$). Then for every $i \in \{1, 2, \ldots, k\}$, we have $d_G(v_i,v_{i+1}) \leq 2$ and $d_G(v_i, v_j) \geq 3$ if $j \notin \{i-1,i+1\}$ and $j \neq i$. For $d_G(v_i,v_{i+1})=2$, let $x_i$ denote a common neighbor of $v_i$ and $v_{i+1}$.

\begin{claim}\label{P4Eedges}
If $\{v_1,v_2,v_3,v_4\}$ induces a $P_4$ in $G^2$ with $d_G(v_i,v_{i+1}) \le 2$ and $v_1v_2 \in E$ then $v_3v_4 \in E$.
\end{claim}

\no{\it Proof of Claim $\ref{P4Eedges}$}:
If $v_1v_2 \in E$ then $v_2v_3 \notin E$ since $d_G(v_1,v_3) \ge 3$. Thus, $d_G(v_2,v_3)=2$; let $x_2$ be a common neighbor of $v_2$ and $v_3$. Now if $d_G(v_3,v_4)=2$ and $x_3$ is a common neighbor of $v_3$ and $v_4$ then, since $\{v_2,v_3,v_4,x_2,x_3\}$ does not induce a bull in $G$, we have $x_2x_3 \notin E$ but then $\{v_1,v_2,x_2,v_3,x_3,v_4\}$ induces a $P_6$ in $G$, which is a contradiction. This shows Claim \ref{P4Eedges}.
$\Diamond$

\medskip

\begin{claim}\label{xixi+1notinE}
For all $i$, if $x_i,x_{i+1},x_{i+2}$ exist, then $x_ix_{i+1} \notin E$ and $x_ix_{i+2} \in E$.
\end{claim}

\medskip

\no{\it Proof of Claim $\ref{xixi+1notinE}$}: Without loss of generality, let $i=1$. Since $\{v_1,x_1,v_2$, $x_2,v_3\}$ does not induce a bull in $G$,
we have $x_1x_2 \notin E$, and thus in general, $x_ix_{i+1} \notin E$. Now since $\{v_1,x_1,v_2,x_2,v_3,x_3\}$ does not induce a $P_6$ in $G$,
we have $x_1x_3 \in E$ and thus in general, $x_ix_{i+2} \in E$. This shows Claim \ref{xixi+1notinE}.
$\Diamond$

\begin{claim}\label{x1x4inE}
For all $i$, if $x_i,x_{i+1},x_{i+3}$ exist, then $x_ix_{i+3} \in E$.
\end{claim}

\medskip

\no{\it Proof of Claim $\ref{x1x4inE}$}: Without loss of generality, let $i=1$. By Claim \ref{P4Eedges}, $x_3$ exists. Then since by Claim \ref{xixi+1notinE} and since $\{v_1, x_1, x_3, v_4, x_4, v_5\}$ does not induce a $P_6$ in $G$, we have $x_1x_4 \in E$, and thus in general, $x_ix_{i+3} \in E$. This shows Claim \ref{x1x4inE}.
$\Diamond$

\medskip

\no {\it Proof of Theorem \ref{GP6bullfredG2holefr} $(i)$}: Suppose to the contrary that $G^2$ contains an even hole $H$ isomorphic to $C_{2k}$, $k \ge 3$. First assume that there is an $i \in \{1,\ldots,2k\}$ with $v_iv_{i+1} \in E$; without loss of generality, say $v_1v_2 \in E$. Then by the distance properties, $d_G(v_2,v_3)=2$, by Claim \ref{P4Eedges}, $v_3v_4 \in E$, and again by the distance properties and by Claim \ref{P4Eedges}, $d_G(v_4,v_5)=2$ and $v_5v_6 \in E$. Now, since $\{v_2,x_2,v_3,v_4,x_4,v_5\}$ does not induce a $P_6$ in $G$, we have $x_2x_4 \in E$ but then $\{v_1,v_2,x_2,x_4,v_5,v_6\}$ induces a $P_6$ in $G$ which is a contradiction.

Thus, for every $i \in \{1,\ldots,2k\}$ $d_G(v_i,v_{i+1})=2$ holds. Clearly, since $\{v_i,x_i,v_{i+1},x_{i+1},v_{i+2},x_{i+2}\}$ does not induce a $P_6$ in $G$, we have $x_ix_{i+2} \in E$ for all $i \in \{1,\ldots,2k\}$. For a $C_6$, this means that $\{x_1,x_3,x_5,v_1,v_6\}$ induces a bull in $G$ which is a contradiction. Now assume that $k \ge 4$. Then by Claim \ref{xixi+1notinE}, $x_1x_3 \in E$ and $x_3x_5 \in E$, and since $\{x_1,x_3,x_5,v_1,v_6\}$ does not induce a bull in $G$, we have $x_1x_5 \notin E$. By Claim \ref{x1x4inE}, we have $x_1x_4 \in E$ and $x_2x_5 \in E$ but now, $\{v_1,x_1,x_4,x_2,x_5,v_6\}$ induces a $P_6$ in $G$ which is a contradiction. This shows that $G^2$ is even-hole-free.

\medskip

Now let $H$ be an odd hole $C_{2k+1}$, $k \ge 2$. First assume that there is an $i \in \{1,\ldots,2k+1\}$ with $v_iv_{i+1} \in E$; without loss of generality, say $v_1v_2 \in E$. Then by the distance properties, $d_G(v_2,v_3)=2$, by Claim \ref{P4Eedges}, $v_3v_4 \in E$, and again by the distance properties and by Claim \ref{P4Eedges}, $d_G(v_4,v_5)=2$ and $v_5v_6 \in E$ and so on, and finally we obtain $v_{2k+1}v_1 \in E$ which is a contradiction to the distance property $d_G(v_{2k+1},v_2) \ge 3$. Thus, for every $i \in \{1,\ldots,2k+1\}$ $d_G(v_i,v_{i+1})=2$ holds. First assume $k \ge 3$. By Claim \ref{x1x4inE}, we have $x_1x_4 \in E$ and $x_2x_5 \in E$ and since $\{x_1,x_3,x_5,v_1,v_6\}$ does not induce a bull in $G$, we have $x_1x_5 \notin E$ but now,
$\{v_1,x_1,x_4,x_2,x_5,v_6\}$ induces a $P_6$ in $G$ which is a contradiction. This shows that $G^2$ is $C_{2k+1}$-free for $k \ge 3$.

\medskip

\no {\it Proof of Theorem \ref{GP6bullfredG2holefr} $(ii)$}: Finally we consider the case when $H$ is a $C_5$ in $G^2$; only in this case we need that $G$ has an e.d. $D$. Recall that for every $i \in \{1,\ldots,5\}$, we have $d_G(v_i,v_{i+1})=2$, $x_ix_{i+1} \notin E$ and $x_ix_{i+2} \in E$.

\begin{claim}\label{notin-D-vertices}
For all $i \in \{1, 2, \ldots, 5\}$, we have $v_i \notin D$ and $x_i \notin D$.
\end{claim}

\no{\it Proof of Claim $\ref{notin-D-vertices}$}: First, without loss of generality, suppose to the contrary that $v_1 \in D$. Then since $D$ is an e.d., $x_1, v_2, x_2, x_4, x_5, v_5 \notin D$. Again, since $D$ is an e.d., there exist $v'_2, v'_5 \in D$ such that $v'_2v_2, v'_5v_5 \in E$. Note that by the distance properties, $v'_2 \neq v'_5$.  Then $\{v'_5, v_5, x_5, v_1, x_1, v_2\}$ induces a $P_6$ in $G$, which is a contradiction. So, for all $i \in \{1,\ldots,5\}$, $v_i \notin D$.

Next, without loss of generality suppose that $x_1 \in D$. Then, since $D$ is an e.d., $x_4, x_5 \notin D$. Since $v_5 \notin D$ and $D$ is an e.d.,
there exists $v'_5 \in D$ such that $v_5 v'_5 \in E$. Then by the distance properties, we see that $\{v'_5, v_5, x_4, x_1, x_3, v_3\}$ induces a $P_6$ in $G$, which is a contradiction. So, for all $i \in \{1,\ldots,5\}$, we have $x_i \notin D$.
$\Diamond$

\medskip

Since $D$ is an e.d. for $G$, for every $i \in \{1,\ldots,5\}$, there exists $v_i' \in D$ such that $v_iv_i' \in E$. Then we have:

\begin{claim} \label{vi'-neq-vj'}
For every $i, j \in \{1, 2, \ldots, 5\}$ with $i \neq j$, $v_i' \neq v_j'$ holds.
\end{claim}

\no{\it Proof of Claim $\ref{vi'-neq-vj'}$}: Suppose to the contrary that $v'_1=v'_2$; clearly $v'_1 \neq v'_3$, and by $v'_1=v'_2$, $v'_1 \neq v'_5$ holds.
 Since $\{v_1,v'_1,v_2,x_2,v_3,v'_3\}$ does not induce a $P_6$ in $G$, we have $v'_1x_2 \in E$ or $v'_3x_2 \in E$.
 Since $\{v_2,v'_1,v_1,x_5,v_5,v'_5\}$ does not induce a $P_6$ in $G$, we have $v'_1x_5 \in E$ or $v'_5x_5 \in E$. Since $\{v'_1,x_2,x_5,v_3,v_5\}$ does not induce a bull in $G$, we have $v'_1x_2 \notin E$ or $v'_1x_5 \notin E$; without loss of generality, assume that $v'_1x_2 \notin E$ holds. This implies $v'_3x_2 \in E$, and since $\{v_2,x_2,v'_3,v_3,x_3\}$ does not induce a bull in $G$, we have $v'_3x_3 \in E$ but now
$\{v_1,v'_1,v_2,x_2,v'_3,x_3\}$ induces a $P_6$ in $G$, which is a contradiction and thus, Claim \ref{vi'-neq-vj'} is shown.
$\Diamond$

\medskip

Now, if $v'_1x_1 \in E$ and $v'_1x_5 \in E$ then $\{v'_2,v_2,x_1,v_1,x_5,v_5\}$ induces a $P_6$ in $G$. Thus, without loss of generality, let us assume that $v'_1x_1 \notin E$ holds.

\no Since $\{v'_1,v_1,x_1,v_2,x_2,v_3\}$ does not induce a $P_6$ in $G$, we have $v'_1x_2 \in E$.

\no Since $\{v'_1,x_2,x_5,v_3,v_5\}$ does not induce a bull in $G$, we have $v'_1x_5 \notin E$.

\no Since $\{v'_1,x_2,x_4,v_3,v_4\}$ does not induce a bull in $G$, we have $v'_1x_4 \notin E$.

Now, $\{v'_1,v_1,x_5,v_5,x_4,v_4\}$ induces a $P_6$ in $G$, which is a contradiction and thus, Theorem \ref{GP6bullfredG2holefr} is shown.
\hfill{$\Box$}

\medskip

Note that Theorem \ref{GP6bullfredG2holefr} implies that the square $G^2$ of any $(P_6$,bull$)$-free graph $G$ with e.d. is hole-free.

\subsection{Squares of ($P_6$, bull)-free graphs with e.d. are banner-free}

\begin{thm}\label{P6-bull-free-ed-implies-banner-free}
Let $G =(V,E)$ be a $(P_6$, bull$)$-free graph. If $G$ has an efficient dominating set, then $G^2$ is banner-free.
\end{thm}

\no{\it Proof.}
Let $G$ be a ($P_6$, bull)-free graph having an efficient dominating set $D$, and suppose to the contrary that $G^2$ contains an induced
banner with vertices $\{v_1, v_2, v_3, v_4, v_5\}$ such that $\{v_1, v_2, v_3, v_4\}$ form a $C_4$ in $G^2$ with edges $v_iv_{i+1}, v_3 v_5 \in E^2$, where $i \in \{1,2,3,4\}$ (index arithmetic modulo 4). Then for $i \in \{1,2,3,4\}$, $d_G(v_i, v_{i+1}) \leq 2$ and $d_G(v_3,v_5) \le 2$, while $d_G(v_1, v_3) \geq 3, ~d_G(v_1, v_5) \geq 3, d_G(v_2, v_4) $ $\geq 3, ~d_G(v_2, v_5) \geq 3$, and $d_G(v_4, v_5) \geq 3$.

If $d_G(v_i,v_{i+1})=2$ for $i \in \{1,2,3,4\}$, then let $x_i$ denote a common neighbor of $v_i$ and $v_{i+1}$. Moreover, if $d_G(v_3,v_5)=2$ then let $y$ denote a common neighbor of $v_3$ and $v_5$; we call $x_i$ and $y$ {\em auxiliary vertices}. By the distance properties, we have $x_iv_j \notin E$ if $j \notin \{i,i+1\}$ and $v_iy \notin E$ for $i \neq 3, i \neq 5$.
Since $G$ is bull-free, $x_ix_{i+1} \notin E$ for $i \in \{1,2,3,4\}$ and $x_2y \notin E$, $x_3y \notin E$ holds.

\begin{claim}\label{v3v5dist2}
$d_G(v_3,v_5)=2$.
\end{claim}

\no{\it Proof of Claim $\ref{v3v5dist2}$}: Suppose to the contrary that $v_3v_5 \in E$. Since $d_G(v_2,v_5) \geq 3$, we have $d_G(v_2,v_3) = 2$, and,
since $d_G(v_4, v_5) \geq 3$, we have $d_G(v_3, v_4) $ $= 2$. So, there exist auxiliary vertices $x_2$ and $x_3$.
Since $d_G(v_2,v_4) \geq 3$, we have $d_G(v_1, v_4)=2$ or $d_G(v_1, v_2)=2$; without loss of generality, let $d_G(v_1, v_2) = 2$. Hence, there exists $x_1$.
Recall that $x_1x_2 \notin E$ since $G$ is bull-free but now $\{v_5, v_3, x_2, v_2, x_1, v_1\}$ induces a $P_6$ in $G$, which is a contradiction.
This shows Claim \ref{v3v5dist2}.
$\diamond$

\medskip

Hence, $d_G(v_3,v_5) = 2$ and the auxiliary vertex $y$ exists. Since $d_G(v_2,v_4) \geq 3$, $d_G(v_2,v_3)=2$ or $d_G(v_3,v_4)=2$ holds.
We show:
\begin{claim}\label{v2v3andv3v4dist2}
$d_G(v_2,v_3)=d_G(v_3,v_4)=2$.
\end{claim}

\no{\it Proof of Claim $\ref{v2v3andv3v4dist2}$}: Without loss of generality, suppose to the contrary that $v_2v_3 \in E$. Hence $d_G(v_3,v_4)=2$ and $x_3$ exists.
 Recall that $x_3y \notin E$ since $G$ is bull-free.
Then, since $\{v_5, y, v_3, x_3, v_4, v_1\}$ does not induce a $P_6$ in $G$, we have $d_G(v_4, v_1) = 2$ and thus, $x_4$ exists.

Moreover, since $d_G(v_1, v_3) \geq 3$, we have $d_G(v_1,v_2)=2$ and $x_1$ exists. Recall that $x_1x_4 \notin E$ since $G$ is bull-free.
Then, by the distance properties, $\{v_4, x_4, v_1, x_1, v_2, v_3\}$ induces a $P_6$ in $G$, which is a contradiction.
By symmetric arguments, we can exclude the case $v_3v_4 \in E$.

This shows Claim \ref{v2v3andv3v4dist2}.
$\diamond$

\medskip

\begin{figure}[t]
\centering
 \includegraphics{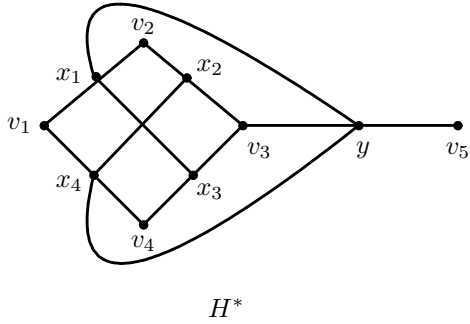}
\caption{The graph $H^*$ used in Theorem \ref{P6-bull-free-ed-implies-banner-free}.}
\label{H3}
\end{figure}

Hence, the auxiliary vertices $x_2$ and $x_3$ exist. Then since $G$ is $P_6$-free, we easily see that $d_G(v_1, v_2) = 2$ and $d_G(v_1, v_4) = 2$.  So, there exist $x_1$ and $x_4$.
Recall that $x_ix_{i+1} \notin E$ for $i \in \{1,2,3,4\}$ and $x_2y \notin E$, $x_3y \notin E$ since $G$ is bull-free.

Then $x_1x_3 \in E$ and $x_2x_4 \in E$ since otherwise, either $\{v_2, x_1, v_1, x_4, v_4, x_3\}$ or  $\{v_4, x_4, v_1, x_2, v_2, x_2\}$ induces a $P_6$ in $G$, which is a contradiction,
and $yx_1, yx_4 \in E$ since otherwise, either $\{y, v_3, x_3, v_4, x_4, v_1\}$ or $\{y, v_3, x_2, v_2, x_1, v_1\}$ induces a $P_6$ in $G$, which is a contradiction.

\medskip

Hence, $G$ contains $H^*$  (see Figure \ref{H3}) as an induced subgraph.

\begin{claim}\label{vertices-notin-D}
$x_1, x_2, x_3, x_4, v_2, v_4 \notin D$.
\end{claim}

\no{\it Proof of Claim $\ref{vertices-notin-D}$}:

\begin{enumerate}
\item[$(i)$] Suppose to the contrary that $x_3 \in D$. Then since $D$ is an e.d., we have $x_1, v_2, x_2 \notin D$. So, there exists $v_2' \in D$ such that $v_2v_2' \in E$ and $v'_2 \neq x_1,x_2$. Then, since $\{v_2', v_2, x_2, v_3, x_3, v_4\}$ does not induce a $P_6$ in $G$, we have $v_2'x_2 \in E$ but then $\{v_2', v_2, x_2, v_3, x_1\}$ induces a bull in $G$, which is a contradiction. Hence, $x_3 \notin D$, and similarly, $x_2 \notin D$.

\item[$(ii)$] Suppose to the contrary that $x_4 \in D$. Then since $D$ is an e.d., we have $y, v_5 \notin D$. So, there exists $v_5' \in D$ such that $v_5v_5' \in E$ and $'v_5 \neq y$. Then since $D$ is an e.d. and by using the distance properties, $\{v_5', v_5, y, x_4, x_2, v_2\}$ induces a $P_6$ in $G$, which is a contradiction. Hence, $x_4 \notin D$, and similarly, $x_1 \notin D$.

\item[$(iii)$] Suppose to the contrary that $v_4 \in D$. Then since $D$ is an e.d., $v_1 \notin D$, and by (ii), $x_1, x_4 \notin D$. Thus, there exists $v_1' \in D$ such that $v_1v_1' \in E$ and $v'_1 \neq x_1,x_4$. Since $d_G(v_1, v_3) \geq 3$, we have $v_1'v_3 \notin E$. So, since $D$ is an e.d.,
$\{v_1', v_1, x_4, v_4, x_3, v_3\}$ induces a $P_6$ in $G$, which is a contradiction. Hence, $v_4 \notin D$, and similarly, $v_2 \notin D$.
\end{enumerate}

Thus, Claim $\ref{vertices-notin-D}$ is proved.
$\Diamond$

\medskip

Since by Claim \ref{vertices-notin-D}, $x_3, x_4, v_4 \notin D$, there exists $v_4' \in D$ such that $v_4v_4' \in E$ and $v'_4 \neq x_3,x_4$. Similarly,
since by Claim \ref{vertices-notin-D}, $x_1, x_2, v_2 \notin D$, there exists $v_2' \in D$ such that $v_2v_2' \in E$ and $v'_2 \neq x_1,x_2$.
Moreover, since $d_G(v_2, v_4) \geq 3$, we have $v_4' \neq v_2'$.

\begin{claim}\label{v2'v3-v4'v3-notin-E}
$v_2'v_3 \notin E$ and $v_4'v_3 \notin E$.
\end{claim}

\no{\it Proof of Claim $\ref{v2'v3-v4'v3-notin-E}$}: Without loss of generality, suppose to the contrary that $v_4'v_3 \in E$.
Since $\{v_4, v_4', v_3, x_2, v_2, v_2'\}$ does not induce a $P_6$ in $G$, we have $v'_2x_2 \in E$ or $v'_4x_2 \in E$.

First assume that $v'_2x_2 \in E$. Then, since $\{x_1,v_2,x_2,v_3,v'_2\}$ does not induce a bull in $G$, we have $v'_2x_1 \in E$. But now
$\{v_4, v_4', v_3, x_2, v'_2, x_1\}$ induces a $P_6$ in $G$, which is a contradiction.

Now assume that $v'_4x_2 \in E$. Then since $D$ is an e.d., $\{v_4', v_4, v_3, x_2, v_2\}$ induces a bull in $G$, which is a contradiction.


Thus, $v'_4v_3 \notin E$ and similarly, $v'_2v_3 \notin E$ which shows Claim \ref{v2'v3-v4'v3-notin-E}.
$\diamond$

\begin{claim}\label{v4'c-v2'b-notin-E}
$v'_2x_2 \notin E$ and $v'_4x_3 \notin E$.
\end{claim}

\no{\it Proof of Claim $\ref{v4'c-v2'b-notin-E}$}: Without loss of generality, suppose to the contrary that $v_4'x_3 \in E$.
Since $\{v'_4, v_4, x_3, v_3, x_4\}$ does not induce a bull in $G$ and by Claim \ref{v2'v3-v4'v3-notin-E}, we have $v'_4x_4 \in E$.
Since $D$ is an e.d., $x_3, v_3, y \notin D$, and by Claim \ref{vertices-notin-D}, $x_2 \notin D$. So, there exists $v_3' \in D$ such that $v_3v_3' \in E$.
By Claim \ref{v2'v3-v4'v3-notin-E}, $v_3' \neq v_2'$ and $v_3'\neq v_4'$.  Since $d_G(v_1, v_3) \geq 3$, $v_3'v_1 \notin E$. Now, since $D$ is an e.d., $\{v_3', v_3, x_3, v_3, v_4, x_4, v_1\}$ induces a $P_6$ in $G$, which is a contradiction. Thus, $v_4'x_3 \notin E$ and similarly $v'_2x_2 \notin E$ which shows
Claim \ref{v4'c-v2'b-notin-E}.
$\Diamond$

%


\medskip

Now, since $\{v'_4, v_4, x_3, v_3, y, v_5\}$ does not induce a $P_6$ in $G$ and by Claims \ref{v2'v3-v4'v3-notin-E} and \ref{v4'c-v2'b-notin-E}, we have
$v'_4y \in E$, and similarly, since $\{v'_2, v_2, x_2, v_3, y, v_5\}$ does not induce a $P_6$ in $G$, we have $v'_2y \in E$ which contradicts the fact that $D$ is an e.d. This finally shows Theorem \ref{P6-bull-free-ed-implies-banner-free}.
\hfill{$\Box$}

\subsection{MWIS problem in (hole, banner)-free graphs}

In this section, we show that the MWIS problem can be solved in time $O(n^2m)$ for (hole,banner)-free graphs. To do this, we need the following:

 For a vertex $v \in V(G)$, the {\it neighborhood} $N(v)$ of $v$ is the set $\{u \in V(G) \mid uv \in E(G)\}$, and its {\it closed neighborhood} $N[v]$ is the set $N(v) \cup \{v\}$.  The neighborhood $N(X)$ of a subset $X \subseteq V(G)$ is the set $\{u \in V(G)\setminus X  \mid  u$ $
\mbox{~is adjacent to a vertex of }X\}$, and its closed neighborhood $N[X]$ is the set $N(X)\cup X$.  Given a subgraph $H$ of $G$ and
$v \in V(G)\setminus V(H)$, let $N_H(v)$ denote the set $N(v) \cap V(H)$, and for $X \subseteq V(G)\setminus V(H)$, let $N_H(X)$ denote the set
$N(X) \cap V(H)$.

A vertex $z \in V(G)$ {\it distinguishes} two other vertices $x, y \in V(G)$ if $z$ is adjacent to one of them and nonadjacent to the other.
A set $M\subseteq V(G)$ is a {\it module} in $G$ if no vertex from $V(G) \setminus M$ distinguishes two vertices from $M$.  The {\it
trivial modules} in $G$ are $V(G)$, $\emptyset$, and all one-vertex sets.  A graph $G$ is {\it prime} if it contains only trivial modules.

A {\it clique} in $G$ is a subset of pairwise adjacent vertices in $G$.  A {\it clique separator} (or {\it clique cutset}) in a connected
graph $G$ is a subset $Q$ of vertices in $G$ which induces a complete graph, such that the graph induced by $V(G) \setminus Q$ is disconnected.  A
graph is an {\it atom} if it does not contain a clique separator.

Let $\cal{C}$ be a class of graphs.  A graph $G$ is {\it nearly $\cal{C}$} if for every vertex $v$ in $V(G)$ the graph induced by
$V(G) \setminus N[v]$ is in $\cal{C}$.

We first note that prime banner-free graphs are $K_{2,3}$-free \cite{BraKleLozMos2010}. We also use the following theorems:

\begin{thm}[\cite{LozMil2008}]\label{thm:LM}
Let $\cal{G}$ be a hereditary class of graphs. If the MWIS problem can be solved in time $O(n^p)$ for prime graphs in $\cal{G}$, where $p \geq 1$ is a constant, then the MWIS problem can be solved for graphs in $\cal{G}$ in time $O(n^p + m)$.
\end{thm}

\begin{thm}[\cite{KarMaf2015}]\label{thm:Tar}
Let $\cal C$ be a class of graphs such that MWIS can be solved in time $O(f(n))$ for every graph in $\cal C$ with $n$ vertices.  Then in any
hereditary class of graphs whose atoms are all nearly $\cal C$ the MWIS problem can be solved in time $O(n^2\cdot f(n))$.
\end{thm}

In \cite{BraKleLozMos2010}, it was shown that prime atoms of (hole, banner)-free graphs are nearly chordal. Applying Corollary 9 in \cite{BraHoa2007} which used an approach for solving MWIS by combining prime graphs and atoms, it was claimed in \cite{BraKleLozMos2010} that MWIS is solvable efficiently for (hole, banner)-free graphs. However, Corollary 9 in \cite{BraHoa2007} is not proven (and thus has to be avoided); a correct way would be to show that atoms of prime (hole, banner)-free graphs are nearly chordal (see also \cite{BraGia2015} for an example). This will be done in the proof of Theorem \ref{prime-hole-Banner-free-implies-nearly-chordal}.
Though the proof given here is very similar to that of \cite{BraKleLozMos2010}, we carefully analyze and reprove it so as to apply the known theorems.

\begin{thm}\label{prime-hole-Banner-free-implies-nearly-chordal}
Every atom of a prime $($hole, banner$)$-free graph is nearly chordal.
\end{thm}

\no {\bf Proof}. Let $G$ be a prime $($hole, banner$)$-free graph and let $G'$ be an atom of $G$. We want to show that $G'$ is nearly chordal, so let us suppose to the contrary that there is a vertex $v \in V(G')$ such that $G' \setminus N[v]$ contains an induced $C_4$, say $H$ with vertex set
$\{v_1, v_2, v_3, v_4\}$ and edge set $\{v_1v_2, v_2v_3, v_3v_4, v_4v_1\}$. For $i \in \{1, \ldots, 4\}$, we define the following:
Let $Q$ denote the component of $G \setminus N[H]$ that contains $v$, let $A_i$ denote the set $\{x \in V(G) \setminus V(H): |N_H(x)| = i\}$,
$A_i^+$ denotes the set $\{x \in A_i \mid N(x) \cap Q \neq \emptyset\}$, and $A^+ = A^+_1\cup A_2^+ \cup A_3^+ \cup A^+_4 $.

Note that by the definition of $Q$ and $A^+$, we have $A^+ = N(Q)$.  Hence $A^+$ is a separator between $H$ and $Q$ in $G$. Throughout this proof, we take all the
subscripts of $v_i$ to be modulo $4$.  Then we have the following:

Since $G$ is banner-free, $A_1^+ \cup A_3^+ = \emptyset$, and so $A^+ = A_2^+ \cup A_4^+$, where $A_2^+ = \cup_{i = 1}^4\{x \in A_2 \mid N(x) \cap V(H) = \{v_i, v_{i+1}\}\}$. Since $G$ is $K_{2,3}$-free and (hole, banner)-free, $A_4^+$ is a clique. Moreover, since $G$ is (hole, banner)-free, we see that
\begin{enumerate}
\item[(1)] $A_2^+$ is a clique, and
\item[(2)] every vertex in $A_2^+$ is adjacent to every vertex in $A_4^+$.
\end{enumerate}

So, $A^+$ is a clique. Since $A^+$ is a separator between $H$ and $Q$ in $G$, we obtain that $V(G')\cap A^+$ is a clique separator in $G'$ between $H$ and
$V(G')\cap Q$ (which contains $v$).  This contradicts the assumption that $G'$ is an atom in $G$.
\hfill{$\Box$}

\medskip

Using Theorem \ref{prime-hole-Banner-free-implies-nearly-chordal}, we now prove the following:

\begin{thm}\label{hole-banner-mwis-time}
 The MWIS problem can be solved in time $O(n^2m)$ for $($hole, banner$)$-free graphs.
\end{thm}

\no{\bf Proof}. Let $G$ be an (hole, banner)-free graph.  First suppose that $G$ is prime. By Theorem~\ref{prime-hole-Banner-free-implies-nearly-chordal}, every atom of $G$ is nearly chordal. Since the MWIS problem can be solved in time $O(m)$ for chordal graphs \cite{Frank1976}, MWIS can be solved in time $O(n^2m)$ for $G$, by Theorem~\ref{thm:Tar}. Then the time complexity is the same when $G$ is not prime, by Theorem \ref{thm:LM}.
\hfill $\Box$

\begin{thm}\label{EDS-P6-bull-free}
 The {\textsc{Min-WED/Max-WED}} can be solved in time $O(n^2m)$ for $(P_6$, bull$)$-free graphs.
 \end{thm}

 \no{\it Proof of Theorem $\ref{EDS-P6-bull-free}$}: Since by Theorem \ref{hole-banner-mwis-time}, the MWIS problem for (hole, banner)-free graphs can be solved in  time $O(n^2m)$, Theorem \ref{EDS-P6-bull-free} follows by Theorems \ref{GP6bullfredG2holefr} and \ref{P6-bull-free-ed-implies-banner-free} and
Theorem \ref{EDS-MWIS-reduction}.
\hfill{$\Box$}

\begin{footnotesize}

\end{footnotesize}


\begin{thebibliography}{99}
\bibitem{Biggs1973}
N. Biggs,
Perfect codes in graphs,
{\sl Journal of Combinatorial Theory, Series B} 15 (1973) 289-296.

\bibitem{BraEscFri2015}
A. Brandst\"adt, E.M. Eschen, and E. Friese,
Efficient domination for some subclasses of $P_6$-free graphs in polynomial time,
CORR arXiv 1503.00091v1, 2015.

\bibitem{BraFicLeiMil2015}
A. Brandst\"adt, P. Fi\v{c}ur, A. Leitert, and  M. Milani\v{c},
Polynomial-time algorithms for weighted efficient domination problems in AT-free graphs and dually chordal graphs,
{\sl Information Processing Letters} 115 (2015) 256--262.

\bibitem{BraGia2015}
A. Brandst\"{a}dt and V. Giakoumakis,
Addendum to: Maximum weight independent sets in hole- and co-chair-free graphs,
{\sl Information Processing Letters} 115 (2) (2015) 345-350.

\bibitem{BraHoa2007}
A. Brandst\"{a}dt and C.T. Ho\'{a}ng,
On clique separators, nearly chordal graphs, and the maximum weight stable set problem,
{\sl Theoretical Computer Science} 389 (2007) 295-306.

\bibitem{BraKleLozMos2010}
A. Brandst\"{a}dt, T. Klembt, V. V. Lozin, and R. Mosca,
On independent vertex sets in subclasses of apple-free graphs,
{\sl Algorithmica} 56 (2010) 383-393.

\bibitem{BraLe2014}
A. Brandst\"adt and V.B. Le,
A note on efficient domination in a superclass of $P_5$-free graphs,
{\sl Information Processing Letters} 114 (2014) 357--359.

\bibitem{BraLeSpi1999}
A. Brandst\"{a}dt, V.B. Le, and J.P. Spinrad,
Graph Classes: A Survey.
SIAM Monographs on Discrete Mathematics, Vol. 3, SIAM, Philadelphia (1999).

\bibitem{BraMilNev2013}
A. Brandst\"adt, M. Milani\v c, and R. Nevries,
New polynomial cases of the weighted efficient domination problem,
{\sl Lecture Notes in Computer Science} 8087 (2013) 195-206.

\bibitem{Chang1997}
M.S. Chang,
Weighted domination of co-comparability graphs,
{\sl Discrete Applied Mathematics} 80 (1997) 135-148.

\bibitem{ChaLiu1993}
M.S. Chang and Y.C. Liu,
Polynomial algorithms for the weighted perfect domination problems on chordal graphs and split graphs,
{\sl Information Processing Letters} 48 (1993) 205-210.

\bibitem{ChaLiu1994}
M.S. Chang and Y.C. Liu,
Polynomial algorithms for the weighted perfect domination problems on interval and circular-arc graphs,
{\sl Journal of Information Sciences and Engineering} 11 (1994) 215-222.

\bibitem{ChaPanCoo1995}
G.J. Chang, C. Pandu Rangan and S.R. Coorg,
Weighted independent perfect domination on co-comparability graphs,
{\sl Discrete Applied Mathematics} 63 (1995) 215-222.

\bibitem{FelHoo1991}
M.R. Fellows and M.N. Hoover,
Perfect domination,
{\sl Australasian Journal of Combinatorics} 3 (1991) 141-150.

\bibitem{Frank1976}
A. Frank,
Some polynomial algorithms for certain graphs and hypergraphs,
in: Proc. of the Fifth British Comb. Conf., Congressus Numerantium, XV (1976) 211-226.

\bibitem{GroLovSch1981}
M. Gr\"otschel, L. Lov\'asz, and A. Schrijver,
The ellipsoid method and its consequences in combinatorial optimization,
{\sl Combinatorica} 1 (1981) 169-197, Corrigendum: {\sl Combinatorica} 4 (1984) 291-295.

\bibitem{HayHedSla1998}
T.W. Haynes, S.T. Hedetniemi, and P.J. Slater,
Fundamentals of Domination in Graphs, Marcel Dekker, New York, 1998.

\bibitem{Karth2015}
T. Karthick,
New polynomial case for efficient domination in $P_6$-free graphs,
{\sl Lecture Notes in Computer Science} 8959 (2015) 81-88.

\bibitem{KarMaf2015}
T. Karthick and F. Maffray,
Maximum weight independent sets in classes related to claw-free graphs,
{\sl Discrete Applied Mathematics} (2015), http://dx.doi.org/10.1016/j.dam.2015.02.012.


\bibitem{LivSto1988}
M. Livingston and Q. Stout,
Distributing resources in hypercube computers,
in: Proceedings of Third Conference on Hypercube Concurrent Computers and Applications (1988) 222-231.

\bibitem{LozMil2008}
V.V. Lozin and M. Milani\v{c},
A polynomial algorithm to find an independent set of maximum weight in a fork-free graph,
{\sl Journal of Discrete Algorithms} 6 (2008) 595-604.

\bibitem{LokVatVil2014}
D. Lokshtanov, M. Vatshelle and Y. Villanger,
Independent set in $P_5$-free graphs in polynomial time,
in: Proceedings of the Twenty-Fifth Annual ACM-SIAM Symposium on Discrete Algorithms (2014) 570-581.

\bibitem{LuTan1998}
C.L. Lu and C.Y. Tang,
Solving the weighted efficient edge domination problem on bipartite permutation graphs,
{\sl Discrete Applied Mathematics} 87 (1998) 203-211.

\bibitem{LuTan2002}
C.L. Lu and C.Y. Tang,
Weighted efficient domination problem on some perfect graphs,
{\sl Discrete Applied Mathematics} 117 (2002) 163-182.

\bibitem{Milan2013}
M. Milani\v c,
Hereditary efficiently dominatable graphs,
{\sl Journal of Graph Theory} 73 (2013) 400-424.

\bibitem{SmaSla1995}
C.B. Smart and P.J. Slater,
Complexity results for closed neighborhood order parameters,
{\sl Congressus Numerantium} 112 (1995) 83-96.

\bibitem{YenLee1996}
C.C. Yen and R.C.T. Lee,
The weighted perfect domination problem and its variants,
{\sl Discrete Applied Mathematics} 66 (1996) 147-160.

\bibitem{Willi2012}
V.V. Williams,
Multiplying matrices faster than Coppersmith–Winograd,
in: Proceedings of the 44th Symposium on Theory of Computing, STOC'12, ACM, New York, USA, 2012, pp.887-898.



\end{thebibliography}
\end{document}